\documentclass[oneside,english]{amsart}
\usepackage[foot]{amsaddr}
\usepackage[T1]{fontenc}
\usepackage[utf8]{inputenc}
\usepackage{lmodern}
\usepackage{amstext}
\usepackage{amsthm}
\usepackage{amssymb}
\usepackage{mathtools}

\makeatletter
\numberwithin{equation}{section}
\numberwithin{figure}{section}

\usepackage{braket}
\usepackage[hidelinks]{hyperref}
\DeclareMathOperator{\Span}{span}

\makeatother

\usepackage{babel}
\begin{document}

\title{Someone shouts, ``$\ket{01000}$!'' Who is Excited?}

\author{Robert S. Smith}
\address{775 Heinz Ave., Berkeley, CA 94710}
\email{\href{mailto:robert@rigetti.com}{robert@rigetti.com}}

\date{November 2, 2017}
\begin{abstract}
We talk about the right way to order amplitudes in a wavefunction,
and the right way to interpret operator matrices in quantum computation.
\end{abstract}

\maketitle
Answer: Both that someone, and qubit $3$.

\section{The Qubit}

A qubit arbitrarily numbered $i$ lives in a space $B_{i}\cong\mathbb{C}^{2}$.
For the purpose of computation, we select two orthonormal vectors
in $B_{i}$, the ground state of qubit $i$ and the excited state,
labeled $\ket{g}_{i}$ and $\ket{e}_{i}$ respectively. Since they're
orthonormal, they form a basis: $B_{i}=\Span\{\ket{g}_{i},\ket{e}_{i}\}$.

Suppose the qubit $i$ is in the notorious \emph{minus state}:
\[
\ket{-}=\frac{1}{\sqrt{2}}\ket{g}_{i}-\frac{1}{\sqrt{2}}\ket{e}_{i}.
\]
When doing algebra, it may be convenient to write it like this, but
sometimes, expressing it as a column vector (otherwise known as a
\emph{wavefunction}) is even more convenient. But which order do we
choose? Should it be
\[
\begin{pmatrix}1/\sqrt{2}\\
-1/\sqrt{2}
\end{pmatrix}\qquad\text{or}\qquad\begin{pmatrix}-1/\sqrt{2}\\
1/\sqrt{2}
\end{pmatrix}\text{?}
\]
The choice is arbitrary; neither has any inherent benefit. Let's choose
the former, because we've been writing $\ket{g}_{i}$ before $\ket{e}_{i}$
in our expressions. In making this choice, we've chosen a \emph{strict
total ordering} on the basis. We can write this as $\ket{g}_{i}\prec\ket{e}_{i}$.

With that, we can canonically assign non-negative integers to the
basis elements, called their \emph{indices}. Consider a finite-dimensional
vector space $V$ which is maybe larger than that of a qubit. With
a basis element $v\in V$, we have
\[
0\leq(\text{the index of \ensuremath{v}})<\dim V
\]
with the property that, with another basis element $u\in V$,
\[
u\prec v\iff(\text{the index of \ensuremath{u}})<(\text{the index of \ensuremath{v}}).
\]

It's important to remind ourselves here that the notion of ``order''
and the associated canonical indices are specific to both the vector
space \emph{and} a choice of basis. Whatever we choose, we call it the \emph{computational basis}.

With indices, we can make our notation a little bit cleaner. Instead
of writing $\ket{g}_i$ or $\ket{e}_i$, we can just identify a basis
element by its index:
\[
\ket{0}_{i}:=\ket{g}_{i}\qquad\text{and}\qquad\ket{1}_{i}:=\ket{e}_{i}.
\]
In the kets, we just write the index as an integer.

Now let's consider more than one qubit. It's well-known that the algebra
of two qubits is best described by the language of tensor products.
That is, for qubits $i$ and $j$, the pair live in the elaborate
space of $B_{i}\otimes B_{j}$. With a choice of basis for $B_{i}$
and $B_{j}$, we have a canonical basis for $B_{i}\otimes B_{j}$,
namely
\[
\big\{\ket{0}_{i}\otimes\ket{1}_{j},\;\ket{0}_{i}\otimes\ket{0}_{j},\;\ket{1}_{i}\otimes\ket{0}_{j},\;\ket{1}_{i}\otimes\ket{1}_{j}\big\}.
\]
But which order should it be in?

If you squint, $\ket{1}_{i}\otimes\ket{0}_{j}$ looks like the number
$10_{2}$, where the subscript $2$ indicates it's written in binary.
This implies the index of $\ket{p}_{i}\otimes\ket{q}_{j}$ should
be $2p+q$, which in turn implies an ordering. Maybe this is too arbitrary
and unappealing, though. So let's turn to an established mathematical
convention related to tensor products.

The Kronecker product of two matrices $A$ and $B$ is written $A\otimes B$
and is defined as the block matrix 
\[
(A\otimes B)_{r,c}:=A_{r,c}B.
\]
The product works just as well for column vectors. Suppose we have
a state of some number of qubits $\ket{\psi}\in V$, identified by
the vector
\[
\begin{pmatrix}\alpha_{0}\\
\alpha_{1}\\
\vdots
\end{pmatrix},
\]
and we wish to throw another qubit to the mix. Maybe this was an ``ambient''
qubit whose state we've insofar ignored. Maybe it's a qubit we are
explicit adding to the system. Let's have qubit $\ket{0}_{\textrm{new}}\in B_{\textrm{new}}$
which is identified by $\begin{psmallmatrix}1\\
0
\end{psmallmatrix}$ . Where do we opt to adjoin this qubit in our space? To the left
or right?

To the left, $\ket{0}_{\textrm{new}}\otimes\ket{\psi}$ looks like
\[
\begin{pmatrix}\alpha_{0}\\
\alpha_{1}\\
\vdots\\
0\\
\vdots
\end{pmatrix},
\]
where the first $\dim V$ entries are the $\alpha$'s and the second
$\dim V$ entries are zero. To the right, $\ket{\psi}\otimes\ket{0}_{\textrm{new}}$
looks like
\[
\begin{pmatrix}\alpha_{0}\\
0\\
\alpha_{1}\\
0\\
\alpha_{2}\\
0\\
\vdots
\end{pmatrix}.
\]
Here, the entries alternate between $\alpha$'s and zeros.

Already, there's a clear winner. Adjoining to the left didn't perturb
the position of our vector entries\footnote{In fact, this would also allow us to manageably work with infinite
tensor products. Consider the infinite tensor product of qubit spaces
$\cdots\otimes B_{2}\otimes B_{1}\otimes B_{0}$. If we say that a
state of this system consists of any tensor product containing finitely
many non-$\ket{0}$ factors, then we can do both linear algebra \emph{and}
true bonafide computation with it by knowing that in the column vector
representation, there will always be a position after
which the entries are zero. Without resorting to trickery, this would
not be so if the infinite product grew to the right.} from $V$ to $B_{\textrm{new}}\otimes V$. The amplitude $\alpha_{j}$
is still at position $j$, now associated with the basis vector $\ket{0}_{\textrm{new}}\otimes\ket{j}_{V}$.
With this, we see that in the space $B_{i}\otimes B_{j}$, the basis
element $\ket{p}_{i}\otimes\ket{q}_{j}$ is identified by the index
$2p+q$, which is in alignment from our earlier observation.

Are there any reasons to do it the other way? The only one I can think
of is that $\bigotimes_{i=0}^{n-1}B_{i}$ is easy to write.

So, it is decided,
\[
\ket{0}_{i}\otimes\ket{0}_{j}\prec\ket{0}_{i}\otimes\ket{1}_{j}\prec\ket{1}_{i}\otimes\ket{0}_{j}\prec\ket{1}_{i}\otimes\ket{1}_{j}.
\]
Since we have an ordering, we have canonical index assignments:
\[
\ket{0}\prec\ket{1}\prec\ket{2}\prec\ket{3}.
\]
Except in general expressions where we're doing arithmetic with the
index, such as writing something like $\ket{2^{k}}$, we might as
well write the indices in binary:
\[
\ket{00}\prec\ket{01}\prec\ket{10}\prec\ket{11}.
\]

Of course, all of this generalizes. Adding an additional qubit will
give us
\[
\ket{000}\prec\ket{001}\prec\ket{010}\prec\ket{011}\prec\ket{100}\prec\ket{101}\prec\ket{110}\prec\ket{111}.
\]
The general rule is thus: For finite-dimensional vector spaces $V$
and $W$ with selected basis elements $\ket{p}_{V}$ and $\ket{q}_{W}$
from the spaces respectively, the induced canonical basis element
of $V\otimes W$ is 
\[
\ket{q+p\dim W}_{V\otimes W}=\ket{p}_{V}\otimes\ket{q}_{W}.
\]

\section{The Operator}

For the rest of this note, it seems natural enough to suppose that
we are working in an $n$-qubit space 
\[
Q_{n}:=B_{n-1}\otimes\cdots\otimes B_{1}\otimes B_{0}.
\]
We wish to consider operators on different subspaces of $Q_{n}$.
Most of the work is done for us; we know what the basis of $Q_{2}$
looks like and we know what the ordering of those elements are. For
instance, the usual $\mathsf{CNOT}:Q_{2}\to Q_{2}$ operator is identified
by the matrix
\[
\begin{pmatrix}
1 & 0 & 0 & 0\\
0 & 1 & 0 & 0\\
0 & 0 & 0 & 1\\
0 & 0 & 1 & 0
\end{pmatrix}
\]
According to our basis ordering, the qubit of $B_{1}$ is the \emph{control
qubit} and the qubit of $B_{0}$ is the \emph{target qubit}. Note
that if we wrote same operator instead as acting on the space $B_{0}\otimes B_{1}$,
it would be
\[
\begin{pmatrix}
1 & 0 & 0 & 0\\
0 & 0 & 0 & 1\\
0 & 0 & 1 & 0\\
0 & 1 & 0 & 0
\end{pmatrix},
\]
in effect transposing the identification of $\ket{01}$ and $\ket{10}$,
causing the middle two rows and middle two columns of the matrix to
be interchanged.

In Quil\footnote{R.\ S.\ Smith, M.\ J.\ Curtis, and W.\ J.\ Zeng, ``A Practical Quantum Instruction Set Architecture,'' (2016), 
  \href{https://arxiv.org/abs/1608.03355}{arXiv:1608.03355}.}, when we write \verb|CNOT 1 3|, what do we mean as an operator
of $Q_{n}$? The following recipe is a constructive way to interpret
it.
\begin{enumerate}
\item Interpret the standard matrix for $\mathsf{CNOT}$ as one written
in the canonical basis for $B_{1}\otimes B_{3}$. That is, the rows
and the columns of the matrix follow the canonical ordering
\[
\ket{0}_{1}\otimes\ket{0}_{3}\prec\ket{0}_{1}\otimes\ket{1}_{3}\prec\ket{1}_{1}\otimes\ket{0}_{3}\prec\ket{1}_{1}\otimes\ket{1}_{3}.
\]
Let's write this operator as $\mathsf{CNOT}_{1,3}$, even though it
is identified by the usual matrix.
\item Lift this operator to a space which is nearly like $Q_{n}$ by tensor
multiplying identities. With $\mathsf{I}_{k}$ the identity operator
on $B_{k}$, write
\[
\mathsf{I}_{n-1}\otimes\cdots\otimes\mathsf{I}_{4}\otimes\mathsf{I}_{2}\otimes\mathsf{I}_{0}\otimes\mathsf{CNOT}_{1,3}.
\]
Note that this lifted operator is identified by a matrix generated
by doing Kronecker multiplication\footnote{This is a subtle point that we wish to distinguish. The operator here
is a linear transformation represented as a tensor product. We can
write a particular representation of this operator by calculating
the following Kronecker products
\[
\begin{psmallmatrix}
1 & 0\\
0 & 1
\end{psmallmatrix}
\otimes\cdots\otimes
\begin{psmallmatrix}
1 & 0\\
0 & 1
\end{psmallmatrix}
\otimes
\begin{psmallmatrix}
1 & 0\\
0 & 1
\end{psmallmatrix}
\otimes
\begin{psmallmatrix}
1 & 0\\
0 & 1
\end{psmallmatrix}
\otimes
\begin{psmallmatrix}
1 & 0 & 0 & 0\\
0 & 1 & 0 & 0\\
0 & 0 & 0 & 1\\
0 & 0 & 1 & 0
\end{psmallmatrix}.
\]
The tensor product results in an automorphism on the space
\[
B_{n-1}\otimes\cdots\otimes B_{4}\otimes B_{2}\otimes B_{0}\otimes B_{1}\otimes B_{3}
\]
while the Kronecker product results in a \emph{representation} of
this operator in the basis we've described in detail.}.
\item Permute the basis vectors so that they match the basis of $Q_{n}$.
This is done with a series of with a transposing isomorphisms $\tau_{i}$,
which are maps swapping the $i$th and $(i+1)$th factors of
the tensor product, from the right:
\[
\tau_{i}:\cdots\otimes V\otimes U\otimes\underbrace{\cdots}_{\mathclap{\text{\ensuremath{i} factors}}}\to\cdots\otimes U\otimes V\otimes\cdots.
\]
In this case, we want to permute
\[
B_{n-1}\otimes\cdots\otimes B_{4}\otimes B_{2}\otimes B_{0}\otimes B_{1}\otimes B_{3}
\]
into order, which will require $\tau_{2}\circ\tau_{1}\circ\tau_{0}\circ\tau_{1}$.
Transpositions have a nice matrix representation, namely
\[
\tau_{i}=
\begin{pmatrix}
1 & 0\\
0 & 1
\end{pmatrix}^{\otimes n-i-2}
\otimes
\begin{pmatrix}
1 & 0 & 0 & 0\\
0 & 0 & 1 & 0\\
0 & 1 & 0 & 0\\
0 & 0 & 0 & 1
\end{pmatrix}
\otimes
\begin{pmatrix}
1 & 0\\
0 & 1
\end{pmatrix}^{\otimes i}.
\]
As we saw with $\mathsf{CNOT}$ before, the matrix in the middle\footnote{This matrix is actually the quantum $\textsf{SWAP}$ operator when
interpreted in $Q_{2}$. While Quil takes care of this whole tranposition
business automatically, we can emulate the effect of $\tau_i$ with the Quil instruction
\[
\texttt{SWAP }(i+1)\texttt{ }i.
\]
} has the effect of reinterpreting basis vectors.
\end{enumerate}
Of course, all of this busywork can be done efficiently.
\end{document}